\documentclass[aps,prl,twocolumn,preprintnumbers,floatfix,nofootinbib,notitlepage,showkeys,showpacs]{revtex4-1}

\usepackage{graphicx}
\usepackage{amssymb}
\usepackage{amsmath}

\newcommand{\adj}{\operatorname{adj}}
\newcommand{\tr}{\operatorname{tr}}
\newcommand{\diag}{\operatorname{diag}}
\newcommand{\res}{\operatorname{res}}

\makeatletter
\newcommand{\xLeftrightarrow}[2][]{\ext@arrow 0099\Leftrightarrowfill@{#1}{#2}}
\makeatother

\begin{document}

\title{Geometry of Flat Directions in Scale-Invariant Potentials}

\author{Kristjan Kannike}
\affiliation{National Institute of Chemical Physics and Biophysics, R\"{a}vala 10, Tallinn 10143, Estonia}

\author{Aleksei Kubarski}
\affiliation{National Institute of Chemical Physics and Biophysics, R\"{a}vala 10, Tallinn 10143, Estonia}

\author{Luca Marzola}
\affiliation{National Institute of Chemical Physics and Biophysics, R\"{a}vala 10, Tallinn 10143, Estonia}

\begin{abstract}
We observe that biquadratic potentials admit non-trivial flat directions when the determinant of the quartic coupling matrix of the scalar fields vanishes. This consideration suggests a new approach to the problem of finding flat directions in scale-invariant theories, noticeably simplifying the study of scalar potentials involving many fields. The method generalizes to arbitrary quartic potentials by requiring that the hyperdeterminant of the tensor of scalar couplings be zero. 
 We demonstrate our approach with detailed examples pertaining to common scalar extensions of the Standard Model.
\end{abstract}

\maketitle

\section{Introduction}

During the last decades, scale-invariant models have taken the spotlight as possible solutions to fundamental issues such as the hierarchy problem \cite{Bardeen:1995kv}, inflation \cite{Albrecht:1982wi,Ellis:1982dg,Ellis:1982ws,Linde:1981mu,Linde:1982zj} and cosmological gravitational wave background \cite{Espinosa:2008kw,Randall:2006py}. 

In this Letter we investigate the appearance of flat directions in scale-invariant scalar potentials, which, in the Gildener-Weinberg approach \cite{Gildener:1976ih}, ensures a successful application of the Coleman-Weinberg (CW) mechanism \cite{Coleman:1973jx} for the radiative generation of mass scales.\footnote{For multi-scale alternatives to the Gildener-Weinberg approach see \cite{Bando:1992wy,Einhorn:1983fc,Ford:1994dt,Ford:1996hd,Ford:1996yc,Casas:1998cf,Chataignier:2018aud}.} In particular, we offer a new implementation of the Gildener-Weinberg method based on the following observation: \emph{a biquadratic scalar potential admits a flat direction if the determinant of its quartic coupling matrix vanishes}. Our technique noticeably simplifies the study of biquadratic potentials involving multiple scalar fields and allows to identify the orientation of the flat direction in a straightforward manner. For a generic potential, instead, the problem is reduced to that of tensor eigenvalues \cite{Qi20051302,2006math......7648L} previously employed in the study of vacuum stability conditions \cite{Kannike:2016fmd,Ivanov:2018jmz}.

We exemplify the method for a biquadratic two-field potential \cite{Hempfling:1996ht} (see also \cite{Loebbert:2018xsd} and Refs. therein), a biquadratic three-field potential \cite{AlexanderNunneley:2010nw,Karam:2015jta,Marzola:2017jzl} and a general two-field potential \cite{AlexanderNunneley:2010nw}.
 
\section{Gildener-Weinberg Approach}
\label{sec:gw} 

We begin by reviewing the Gildener-Weinberg formalism \cite{Gildener:1976ih}, following the conventions of Ref. \cite{AlexanderNunneley:2010nw}. Consider a general renormalizable gauge theory where the field vector $\mathbf{\Phi}$ contains $n$ real scalar degrees of freedom. A generic quartic potential is of the form  
\begin{equation}
  V(\mathbf{\Phi}) = \frac{1}{4!} \sum_{i,j,k,l} \lambda_{ijkl} \phi_i \phi_j \phi_k \phi_{l},
\label{eq:V:general}
\end{equation}
with $\lambda_{ijkl}$ being the symmetric tensor of quartic couplings. At tree-level all fields are massless and the potential is bounded from below if $V(\mathbf{\Phi}) \geq 0$ for all $\mathbf{\Phi}$.

We parameterize the field as $\mathbf{\Phi} = \varphi \mathbf{N}$, where $\varphi$ is the radial coordinate and $\mathbf{N}$ is a unit vector. The renormalization scale $\mu$ is chosen such that the minimum of the potential is zero on the unit sphere $\mathbf{N}^{T} \mathbf{N} = 1$.  The unit vector $\mathbf{N} = \mathbf{n}$ that realizes the minimum points along the flat direction $\mathbf{\Phi} = \varphi \mathbf{n}$. Because the flat direction is a stationary line, it satisfies $\nabla_{\mathbf{N}} V({\mathbf{N}})|_{\mathbf{N} = \mathbf{n}} = 0$, implying
\begin{equation}
  \sum_{j,k,l} \lambda_{ijkl} n_{j} n_{k} n_{l} = 0.
  \label{eq:cond1}
\end{equation}
If the Hessian matrix
\begin{equation}
   (\mathbf{P})_{ij} = \left. \frac{\partial^{2}{V}(\mathbf{N})}{\partial N_{i} \partial N_{j}}\right|_{\mathbf{N} = \mathbf{n}} = \frac{1}{2} \sum_{k,l} \lambda_{ijkl} n_{k} n_{l}
\label{eq:gw:hessian}
\end{equation}
is positive-semidefinite, the flat direction is a local minimum.

Quantum corrections then shape the potential along $\mathbf{n}$
\begin{equation}
  V^{(1)}(\varphi \mathbf{n}) = A(\mathbf{n}) \varphi^{4} + B(\mathbf{n}) \varphi^{4} \ln \frac{\varphi^{2}}{\mu^{2}},
\label{eq:CW}  
\end{equation}
where the constants given in the $\overline{\text{MS}}$ scheme, at the one-loop level,
\begin{align}
  A(\mathbf{n}) &= \frac{1}{64 \pi^{2} v_{\varphi}^{4}} \left\{ 
  \tr \left[ \mathbf{m}_{S}^{4} \left( \ln \frac{\mathbf{m}_{S}^{2}}{v_{\varphi}^{2}} - \frac{3}{2} \right) \right] 
  \right.
  \notag
  \\
  & + 3 \tr \left[ \mathbf{m}_{V}^{4} \left( \ln \frac{\mathbf{m}_{V}^{2}}{v_{\varphi}^{2}} - \frac{5}{6} \right) \right]
  \notag
  \\
  & \left. \left. 
  - 4 \tr \left[ \mathbf{m}_{F}^{4} \left( \ln \frac{\mathbf{m}_{F}^{2}}{v_{\varphi}^{2}} - \frac{3}{2} \right) \right]
  \right\},
  \right.
  \\
  B(\mathbf{n}) &= \frac{1}{64 \pi^{2} v_{\varphi}^{4}} \left( \tr \mathbf{m}_{S}^{4} + 3 \tr \mathbf{m}_{V}^{4} - 4 \tr \mathbf{m}_{F}^{4} \right)
\end{align}
depend on the tree-level scalar, vector and fermion mass matrices, $\mathbf{m}^{2}_{S,V,F}$, evaluated at the radiatively induced minimum $\mathbf{\Phi}= v_{\varphi} \mathbf{n}$. The tree-level scalar mass matrix is related to the Hessian \eqref{eq:gw:hessian} via
\begin{equation}
  \mathbf{m}_{S}^{2} = v_{\varphi}^{2} (\mathbf{P})_{ij}.
\end{equation}

Minimizing the potential \eqref{eq:CW} with respect to the radial coordinate $\varphi$ shows the presence of a non-trivial stationary point at the renormalization scale $\mu = v_{\varphi} \exp (\frac{A}{2 B} + \frac{1}{4})$, yielding 
\begin{equation}
  V^{(1)}(\varphi \mathbf{n}) = B(\mathbf{n}) \varphi^{4} \left( \ln \frac{\varphi^{2}}{v_{\varphi}^{2}} - \frac{1}{2} \right).
\label{eq:CW:improved}
\end{equation}
The quantum-corrected mass matrix is
\begin{equation}
  (\mathbf{m}_{S}^{2} + \delta\mathbf{m}_{S}^{2})_{ij} = \left. \frac{\partial^{2} [V(\mathbf{\Phi}) + 
  V^{(1)}(\mathbf{\Phi})]}{\partial \phi_{i} \partial \phi_{j}}\right|_{\mathbf{\Phi} = v_{\varphi} \mathbf{n}},
\end{equation}
so the field along the flat direction -- the pseudo-Goldstone of classical scale invariance -- obtains a mass $m_{\varphi}^{2} = 8 B v_{\varphi}^{2}$. The masses of the orthogonal fields receive negligible corrections.

\section{A New Approach to Flat Directions in Scale-Invariant Potentials}
\label{sec:flat}

We present an alternative implementation of the Gildener-Weinberg approach,  demonstrating that a flat direction appears in a biquadratic scalar potential if the determinant of the quartic coupling matrix vanishes. For generic potentials, the flat direction appears when the \emph{hyperdeterminant} \cite{Qi20051302,2006math......7648L} of the quartic coupling tensor vanishes.

\subsection{Biquadratic Potentials}

A generic biquadratic potential of $n$ real scalar fields organized in the vector $\mathbf{\Phi}$ is given by
\begin{equation}
  V(\mathbf{\Phi}) = \sum_{i,j} \phi_i^2 \lambda_{ij} \phi_j^2 
  = (\mathbf{\Phi}^{\circ 2})^T \mathbf{\Lambda} \mathbf{\Phi}^{\circ 2},
\label{eq:biq:pot}
\end{equation}
where $\mathbf{\Lambda}$ is the symmetric matrix of the quartic couplings and the $\circ$ symbol indicates the \emph{Hadamard product}, defined as the element-wise product of matrices of same dimensions: $(\mathbf{A} \circ \mathbf{B})_{ij} = A_{ij} B_{ij}$. The \emph{Hadamard power} is given by $(\mathbf{A}^{\circ n})_{ij} = A_{ij}^{n}$.

The potential is required to be bounded from below, for which a necessary and sufficient condition is that the matrix $\mathbf{\Lambda}$ be copositive \cite{Kannike:2012pe}. This can be ascertained via the Cottle-Habetler-Lemke theorem \cite{Cottle1970295}: Suppose that the order $n-1$ principal submatrices of a real symmetric matrix $\mathbf{A}$ of order $n$ are copositive. Then $\mathbf{A}$ is copositive if and only if 
\begin{equation}
  \det (\mathbf{A}) \geq 0 \quad \lor \quad \text{some element(s) of } \adj \mathbf{A} < 0.
\end{equation}
The adjugate $ \adj(\mathbf{A})$ of a matrix $\mathbf{A}$ is defined through the relation $\mathbf{A} \adj(\mathbf{A}) = \det(\mathbf{A}) \, \mathbf{I}$.
Notice that a semipositive-definite matrix is copositive, therefore a semipositive-definite $\mathbf{\Lambda}$ is sufficient for the stability of the potential.

The norm of $\mathbf{\Phi}$ can be expressed through its Hadamard square as
\begin{equation}
  \mathbf{\Phi}^{T} \mathbf{\Phi} = \mathbf{e}^{T} \mathbf{\Phi}^{\circ 2},
\label{eq:norm}  
\end{equation}
where $\mathbf{e}= (1, \ldots, 1)^{T}$, the vector of ones, is an identity element of the Hadamard product. The relation \eqref{eq:norm} is quadratic in the field $\mathbf{\Phi}$ but only linear in $\mathbf{\Phi}^{\circ 2}$.

We restrict the potential \eqref{eq:biq:pot} to the unit hypersphere by means of a Lagrange multiplier,
\begin{equation}
  V(\mathbf{N}, \lambda) = (\mathbf{N}^{\circ 2})^T \mathbf{\Lambda} \mathbf{N}^{\circ 2}
  + \lambda (1 - \mathbf{e}^{T} \mathbf{N}^{\circ 2}).
\label{eq:biq:pot:hypersph}
\end{equation}
so that its minimization equation recovers
\begin{equation}
\mathbf{e}^T \mathbf{N}^{\circ 2} = 1.
\label{eq:biq:norm:phi}
\end{equation}
While the vector $\mathbf{N}$ lies on the unit hypersphere, its Hadamard square $\mathbf{N}^{\circ 2}$ lies on the \emph{unit simplex} with its elements as barycentric coordinates. Extremizing the potential on the unit hypersphere is therefore equivalent to extremizing a quadratic function of $\mathbf{N}^{\circ 2}$ on the unit simplex. This problem is known within the theory of optimization as \textit{the standard quadratic program} (see e.g. \cite{Fletcher:1987:PMO:39857}).

The vector of first derivatives of the potential is
\begin{equation}
  \nabla_{\mathbf{N}} V = 4 \mathbf{N} \circ \mathbf{\Lambda}\mathbf{N}^{\circ 2} 
  - 2 \lambda \mathbf{N} \circ \mathbf{e}
   = 2 \mathbf{N} \circ ( 2 \mathbf{\Lambda} \mathbf{N}^{\circ 2} - \lambda \mathbf{e} ),
\label{eq:biq:grad}
\end{equation}
and stationary points obey $\nabla_{\mathbf{N}} V = \mathbf{0}$. 

At first, we assume that none of the elements of $\mathbf{N}^{\circ 2}$ vanish. Under this assumption
\begin{equation}
   2  \mathbf{\Lambda}\mathbf{N}^{\circ 2} = \lambda \mathbf{e}.
\label{eq:biq:phi:sq:min:lambda}
\end{equation}
Multiplying both sides of Eq.~\eqref{eq:biq:phi:sq:min:lambda} from the left by $(\mathbf{N}^{\circ 2})^{T}$ and using the constraint \eqref{eq:biq:norm:phi} yields
\begin{equation}
  \lambda = 2 (\mathbf{N}^{\circ 2})^T \mathbf{\Lambda} \mathbf{N}^{\circ 2} = 2 V(\mathbf{N}).
\end{equation}
Eq.~\eqref{eq:biq:phi:sq:min:lambda} then gives  
\begin{equation}
  \mathbf{\Lambda} \mathbf{N}^{\circ 2}
  = \left[(\mathbf{N}^{\circ 2})^T \mathbf{\Lambda} \mathbf{N}^{\circ 2} \right] \mathbf{e}
  \equiv V(\mathbf{N}) \, \mathbf{e},
\label{eq:biq:phi:sq:min:V}
\end{equation}
which we solve by making the ansatz
\begin{equation}
  \mathbf{N}^{\circ 2} = C \adj(\mathbf{\Lambda}) \, \mathbf{e},
\label{eq:biq:ansatz:sol}
\end{equation}
where $C$ is a real normalization constant. Inserting Eq.~\eqref{eq:biq:ansatz:sol} into Eq.~\eqref{eq:biq:phi:sq:min:V}, we obtain
\begin{equation}
  C = \frac{1}{\mathbf{e}^T \! \adj(\mathbf{\Lambda}) \, \mathbf{e}},
 \label{eq:biq:ansatz:norm}
\end{equation}
normalizing $\mathbf{N}$ to unity. Note that $\adj(\mathbf{\Lambda}) \, \mathbf{e}$ is the vector of row sums of $\adj(\mathbf{\Lambda})$, and $\mathbf{e}^T \! \adj(\mathbf{\Lambda}) \, \mathbf{e}$ is the sum of all elements of $\adj(\mathbf{\Lambda})$. Consistency requires the elements of $\mathbf{N}^{\circ 2}$ to be positive and finite\footnote{To this end, it is sufficient, but not necessary, that all the off-diagonal elements of $\mathbf{\Lambda}$ be negative.} to ensure that the one-loop corrections will result in a positive and finite vacuum expectation value.

The value of the potential at the extremum $\mathbf{N}$ on the unit hypersphere is
\begin{equation}
V(\mathbf{N}) = \frac{\det (\mathbf{\Lambda}) }{ \mathbf{e}^T \! \adj (\mathbf{\Lambda}) \mathbf{e}}.
\label{eq:V:on:hypersphere}
\end{equation}
Consequently, the determinant of $\mathbf{\Lambda}$ must be zero to have a flat direction along $\mathbf{N} = \mathbf{n}$:
\begin{equation}
  \det(\mathbf{\Lambda}) = 0 \xLeftrightarrow{0 < \mathbf{n}^{\circ 2} < \infty} V(\mathbf{n}) = 0.
\label{eq:det:0:V:0} 
\end{equation}
The Hadamard square of the unit vector 
\begin{equation}
  \mathbf{n}^{\circ 2} = \frac{\adj(\mathbf{\Lambda}) \, \mathbf{e}}{\mathbf{e}^T \! \adj(\mathbf{\Lambda}) \, \mathbf{e}}
\label{eq:flat:direction:sol}
\end{equation} 
is then an eigenvector of $\mathbf{\Lambda}$ with a null eigenvalue.

For  $\mathbf{n}$ to correspond to a stable flat direction, the solution must be a minimum of the scalar potential on the hypersphere. Using Eq.~\eqref{eq:biq:grad}, the Hessian \eqref{eq:gw:hessian} of $V(\mathbf{N}, \lambda)$ is
\begin{equation}
\begin{split}
   \mathbf{P} &= \nabla_{\mathbf{N}} \nabla_{\mathbf{N}}^{T} V(\mathbf{N}, \lambda)|_{\mathbf{N} = \mathbf{n}}
   \\
   &= \diag \left[ 2 \, (2 \mathbf{\Lambda} \mathbf{n}^{\circ 2} - \lambda \mathbf{e}) \right]
   + 8 \mathbf{\Lambda} \circ (\mathbf{n} \mathbf{n}^{T})
   \\
   &= \diag ( 4 \mathbf{\Lambda} \mathbf{n}^{\circ 2} ) + 8 \mathbf{\Lambda} \circ (\mathbf{n} \mathbf{n}^{T}),
\end{split}
\label{eq:hessian}
   \end{equation}
where $\diag( \mathbf v)$ designates a diagonal matrix whose diagonal is given by the vector $\mathbf v$, and we used $\lambda = 2 V(\mathbf{n}) = 0$. The condition for the extremum on the unit hypersphere to be a local minimum is that the Hessian matrix $\mathbf{P}$ be positive on the tangent space of the hypersphere at $\mathbf{N} = \mathbf{n}$ \cite{avriel2003nonlinear}. Because $\mathbf{n}$ is an eigenvector of $\mathbf{P}$ with zero eigenvalue, and given that any vector in the field space can be written as a linear combination of $\mathbf{n}$ and the tangent vectors, $\mathbf{P}$ must be positive-semidefinite on all unit vectors $\mathbf{N}$.

Along the flat direction, the diagonal term in the expression \eqref{eq:hessian} vanishes and the Hessian is $\mathbf{P} = 8 \mathbf{\Lambda} \circ (\mathbf{n} \mathbf{n}^{T}) = 8 \diag(\mathbf{n})  \mathbf{\Lambda} \diag(\mathbf{n})$. Since $\det (\mathbf{P}) = \det(\mathbf{\Lambda}) \prod 8 n_{i}^{2}$, the Hessian is positive-semidefinite if and only if the coupling matrix $ \mathbf{\Lambda}$ is positive-semidefinite, ensuring also that the potential is bounded from below. As $\det(\mathbf{\Lambda})$ goes to zero, the mass of the radial degree of freedom in the direction of $\mathbf{n}$ vanishes.

In the cases where up to $n-1$ components of the unit vector $\mathbf{N}$ vanish, Eq.~\eqref{eq:biq:phi:sq:min:lambda} is restricted to the coupling matrix for the fields in the non-degenerate subspace. Without loss of generality, the coupling matrix and the flat direction can then be brought in the block form  
\begin{equation}
  \mathbf{\Lambda} = 
  \begin{pmatrix}
    \mathbf{\Lambda}_{11} & \mathbf{\Lambda}_{12}
    \\
    \mathbf{\Lambda}_{12}^{T} &\mathbf{\Lambda}_{22}
  \end{pmatrix},
  \quad
  \mathbf{n} = 
  \begin{pmatrix}
    \mathbf{n}_{1} \\ \mathbf{0}
  \end{pmatrix}
\label{eq:hessian:block}
\end{equation}
by a suitable permutation of the fields.

The square $\mathbf{n}_{1}^{\circ 2}$ of the non-zero subvector is an eigenvector of the submatrix $\mathbf{\Lambda}_{11}$ with the zero eigenvalue and the Hessian \eqref{eq:hessian} takes the form
\begin{equation}
  \mathbf{P} =   \begin{pmatrix}
    8 \mathbf{\Lambda}_{11} \circ (\mathbf{n}_{1} \mathbf{n}_{1}^{T}) & \mathbf{0}
    \\
    \mathbf{0}^{T} &\diag(4 \mathbf{\Lambda}_{12}^{T} \mathbf{n}_{1}^{\circ 2})
  \end{pmatrix}.
\label{eq:hessian:part:zero}
\end{equation}
Thus, $\mathbf{n}$ is a stable flat direction if $\mathbf{\Lambda}_{11}$ is positive-semidefinite and $\mathbf{\Lambda}_{12}^{T} \mathbf{n}_{1}^{\circ 2}$ is a vector with positive elements. The $\mathbf{\Lambda}_{22}$ submatrix must be copositive for the potential to be bounded from below.

Conversely, if the desired field direction $\mathbf{n}$ and particle masses are known, then the coupling matrix $\mathbf{\Lambda}$ can be found by taking the element-wise Hadamard inverse of the relation \eqref{eq:hessian}:
\begin{equation}
  \mathbf{\Lambda} = \frac{1}{8} \mathbf{P} \circ (\mathbf{n} \mathbf{n}^{T})^{\circ -1} = \frac{1}{8 v_{\varphi}^{2}} \mathbf{m}_{S}^{2} \circ (\mathbf{n} \mathbf{n}^{T})^{\circ -1}.
\label{eq:lambda:from:hessian}
\end{equation}
If some elements of $\mathbf{n}$ are zero, the Hadamard inverse in Eq.~\eqref{eq:lambda:from:hessian} has to be restricted to the submatrices $\mathbf{\Lambda}_{11}$ and $\mathbf{n}_{1} \mathbf{n}_{1}^{T}$ in Eq.~\eqref{eq:hessian:block}. The undetermined elements of $\mathbf{\Lambda}$ must be chosen such as to keep the  flat direction $\mathbf{n}$ a minimum as per Eq.~\eqref{eq:hessian:part:zero}.

\subsection{General Potentials}

The generic potential in Eq.~\eqref{eq:V:general}, restricted to the unit hypersphere, can be written as
\begin{equation}
  V(\mathbf{N}, \lambda) = \frac{1}{4} \mathbf{\Lambda} \mathbf{N}^{4}
  + \lambda (1 - \mathbf{N}^{T} \mathbf{N}).
\label{eq:biq:pot:hypersph}
\end{equation}
A polynomial $f(\mathbf{x})$ of order  $m$ and its coefficient tensor $\mathbf{A}$ are related as $\mathbf{A} \mathbf{x}^{m} = m f(\mathbf{x})$ and $\mathbf{A} \mathbf{x}^{m-1} = \nabla_{\mathbf{x}} f(\mathbf{x})$ with
\begin{equation}
  \sum_{i_{1}, i_{2}, \ldots, i_{m}} A_{i_{1}i_{2}\ldots i_{m}} x_{i_{1}} x_{i_{2}} \cdots x_{i_{m}} = \mathbf{x}^{T} \mathbf{A} \mathbf{x}^{m-1}.
\end{equation}

The extrema conditions for the potential \eqref{eq:biq:pot:hypersph} are
\begin{align}
  \mathbf{\Lambda} \mathbf{N}^{3} &= \lambda \mathbf{N},
  \label{eq:Lambda:E:evals}
  \\
  \mathbf{N}^{T} \mathbf{N} &= 1,
  \label{eq:Lambda:E:sph}
\end{align}
which we recognize as the \emph{E-eigenvalue equations} for the $\mathbf{\Lambda}$ tensor \cite{Qi20051302,2006math......7648L}. Tensors also possess N-eigenvectors and N-eigenvalues given by the solutions of $\mathbf{A} \mathbf{x}^{m-1} = \lambda \mathbf{x}^{\circ (m-1)}$. It is the E-eigenvectors, however, which are normalized to the unit hypersphere. We refer the reader to the book in Ref. \cite{doi:10.1137/1.9781611974751} for further details.

The number of E-eigenvectors of a symmetric tensor of order $m$ in $\mathbb{R}^n$ is
\begin{equation}
  d = \frac{(m - 1)^{n} - 1}{m - 2}.
\label{eq:number:eigenvectors}
\end{equation}
In general, they can be complex, but only real tensor eigenvectors and eigenvalues are physical solutions of Eqs.~\eqref{eq:Lambda:E:evals} and \eqref{eq:Lambda:E:sph}.

Eliminating $N_{i}$ from Eqs.~\eqref{eq:Lambda:E:evals} and \eqref{eq:Lambda:E:sph} yields the characteristic polynomial $\phi_{\mathbf{\Lambda}}(\lambda)$ of the tensor, with a degree given by Eq.~\eqref{eq:number:eigenvectors}. The multivariate resultant of a system of polynomial equations is a polynomial in their coefficients, which vanishes if and only if the equations have a common root. The free term of $\phi_{\mathbf{\Lambda}}(\lambda)$ -- the product of all E-eigenvalues -- is given by the resultant $\res_{\mathbf{N}} (\mathbf{\Lambda} \mathbf{N}^{3})$. For that reason, the resultant is also called the hyperdeterminant. In order to have a zero tensor eigenvalue, we must have $\res_{\mathbf{N}} (\mathbf{\Lambda} \mathbf{N}^{3}) = 0$, implying that $\mathbf{\Lambda} \mathbf{N}^{3} = \mathbf{0}$ has a non-trivial solution. 

The tensor $\mathbf{\Lambda}$ must also be positive-semidefinite in order for the potential \eqref{eq:V:general} to be bounded from below: all of its eigenvalues and those of its principal subtensors -- obtained by setting one or more fields to zero in $V(\mathbf{N})$ -- must be non-negative. The Hessian \eqref{eq:gw:hessian} must be positive-semidefinite as well.

Note that for a quartic potential of two fields, the resultant is proportional to its discriminant with one field set to unity. Unfortunately, for a larger number of variables, calculation of the resultant is prohibitively expensive.

Unlike for biquadratic potentials, in general all the potential coefficients cannot be determined from knowledge of the mass matrix.

\section{Examples}
\label{sec:examples}

\subsection{Biquadratic Two-Field Potential}

For a biquadratic potential of two fields
\begin{equation}
  V = \lambda_{1} \phi_{1}^{4} + \lambda_{12} \phi_{1}^{2} \phi_{2}^{2} + \lambda_{2} \phi_{2}^{4},
\label{eq:V:biquadratic:2:fields}
\end{equation}
the coupling matrix and its adjugate are determined via Eq.~\eqref{eq:biq:pot} as 
\begin{equation}
  \mathbf{\Lambda} = 
  \begin{pmatrix}
    \lambda_{1} & \frac{1}{2} \lambda_{12} \\
    \frac{1}{2} \lambda_{12} & \lambda_{2}
  \end{pmatrix},
  \,\,
  \adj(\mathbf{\Lambda})=
	\begin{pmatrix}
	\lambda_2 & -\frac{1}{2}\lambda_{12} \\
	-\frac{1}{2}\lambda_{12} & \lambda_1
  \end{pmatrix}.
\end{equation}

The extrema equations of $V(\mathbf{N}, \lambda)$ on the unit simplex have three solutions, two corresponding to either of the two fields set to zero and a solution where neither vanishes. For the last one, from Eqs.~\eqref{eq:biq:ansatz:sol} and \eqref{eq:biq:ansatz:norm}, we have
\begin{equation}
  \mathbf{N}^{\circ 2} = \frac{1}{\lambda_1 - \lambda_{12} + \lambda_2} 
  \begin{pmatrix}
    \lambda_{2} - \frac{\lambda_{12}}{2} 
    \\
    \lambda_{1} - \frac{\lambda_{12}}{2}
  \end{pmatrix}.
\label{eq:2:fields:N:2}
\end{equation}
The value of the potential along the direction $\mathbf{N}$ is
\begin{equation}
		V(\mathbf{N}) = \frac{4\lambda_1 \lambda_2 - \lambda_{12}^2}{4(\lambda_1 - \lambda_{12} + \lambda_2)}
\end{equation}
and the flat direction $\mathbf{N}=\mathbf{n}$ appears when
\begin{equation}
  4 \det \mathbf{\Lambda} = 4 \lambda_{1} \lambda_{2} - \lambda_{12}^{2} = 0,
\label{eq:2:fields:det:0}
\end{equation}
implying $V(\mathbf{n})=0$. We must have a negative $\lambda_{12} = -2 \sqrt{\lambda_{1} \lambda_{2}}$ in order for both elements of $\mathbf{n}^{\circ 2}$ to be positive.

The Hessian matrix \eqref{eq:hessian}, given by
	\begin{equation}
		\mathbf{P} =
		\begin{pmatrix}
		12\lambda_1 N_1^2+2\lambda_{12} N_2^2 & 4\lambda_{12} N_1 N_2 \\
		4\lambda_{12} N_1 N_2 & 12\lambda_2 N_2^2+2\lambda_{12} N_1^2
		\end{pmatrix},
\end{equation}
is positive-semidefinite if $\mathbf{\Lambda}$ is, requiring that all principal minors of $\mathbf{\Lambda}$ be non-negative: $\det \mathbf{\Lambda} \geq 0$, $\lambda_{1} \geq 0$, $\lambda_{2} \geq 0$.

The two solutions corresponding to vanishing fields are given by $N_{1} = 1$, $N_{2} = 0$ with $V(\mathbf{N}) = \lambda_{1}$, and $N_{1} = 0$, $N_{2} = 1$ with $V(\mathbf{N}) = \lambda_{2}$. The flat direction is then obtained when the self-coupling of the non-zero field vanishes. In the first case, for example, we have $\lambda_{1} = 0$. The block form of the Hessian matrix in Eq.~\eqref{eq:hessian:block} is given by
$\mathbf{\Lambda}_{11} = \begin{pmatrix} \lambda_{1} \end{pmatrix}$, $\mathbf{\Lambda}_{12} = \begin{pmatrix} \lambda_{12} \end{pmatrix}$, $\mathbf{\Lambda}_{22} = \begin{pmatrix} \lambda_{2} \end{pmatrix}$ and $\mathbf{n}_{1} = \begin{pmatrix} 1 \end{pmatrix}$, so the requirement of positive $\diag(\mathbf{\Lambda}_{12}^{T} \mathbf{n}_{1}^{\circ 2}) = \begin{pmatrix} \lambda_{12} \end{pmatrix}$ implies $\lambda_{12} > 0$. The copositivity of $\mathbf{\Lambda}_{22}$ requires $\lambda_{2} > 0$.
 
To relate our formalism to the hyperspherical coordinate approach, notice that the mixing angle between the fields in Eq.~\eqref{eq:2:fields:N:2} is 
	\begin{equation}
		\tan^2 \theta = \frac{n_{2}^{2}}{n_{1}^{2}} = \frac{2\lambda_1-\lambda_{12}}{2\lambda_2 -\lambda_{12}} = \sqrt{\frac{\lambda_{1}}{\lambda_{2}}}.
\end{equation}

In polar coordinates $N_{1} = \cos \theta$ and $N_{2} = \sin \theta$, the same angle is obtained from the conditions $V = dV/d\theta = 0$. The  two other solutions of these equations, $\theta = 0$ and $\theta = \frac{\pi}{2}$, correspond to a flat direction lying on a coordinate axis.

Conversely, let us assume a flat direction at a given angle $\theta$. The mixing matrix is
\begin{equation}
  \mathbf{O} = \begin{pmatrix}
    \cos \theta & -\sin \theta
    \\
    \sin \theta & \cos \theta
  \end{pmatrix}
\end{equation}
and the mass eigenstates $h_1$ and $h_2$, respectively along and orthogonal to the flat direction, are obtained as
\begin{equation}
  \begin{pmatrix}
    h_{1}
    \\
    h_{2}
  \end{pmatrix}
  = \mathbf{O}^{T}
    \begin{pmatrix}
    \phi_{1}
    \\
    \phi_{2}
  \end{pmatrix}.
\end{equation}
At tree-level $M_{1} = 0$, so the scalar mass matrix is 
\begin{equation}
  \mathbf{m}_{S}^{2} = M_{2}^{2}
  \begin{pmatrix}
    \sin^{2} \theta & -\sin \theta \cos \theta 
    \\
    -\sin \theta \cos \theta & \cos^{2} \theta
  \end{pmatrix}.
\end{equation}
The relation~\eqref{eq:lambda:from:hessian} then determines the quartic coupling matrix 
\begin{equation}
  \mathbf{\Lambda} = \frac{1}{8} \frac{M_{2}^{2}}{v_{\varphi}^{2}}
  \begin{pmatrix}
    \tan^{2} \theta & -1 
    \\
    -1 & \cot^{2} \theta
  \end{pmatrix}.
\end{equation}

\subsection{Biquadratic Three-Field Potential}

For a biquadratic potential of three fields
\begin{equation}
\begin{split}
  V &= \lambda_{1} \phi_{1}^{4} + \lambda_{12} \phi_{1}^{2} \phi_{2}^{2} + \lambda_{2} \phi_{2}^{4}
  + \lambda_{23} \phi_{2}^{2} \phi_{3}^{2}
  \\
  &+ \lambda_{3} \phi_{3}^{4} + \lambda_{13} \phi_{1}^{2} \phi_{3}^{2},
\end{split}
\label{eq:V:biquadratic:2:fields}
\end{equation}
the coupling matrix and its adjugate are given by
\begin{widetext}
	\begin{equation}
    \mathbf{\Lambda} =     
  \begin{pmatrix}
    \lambda_{1} & \frac{1}{2} \lambda_{12} & \frac{1}{2} \lambda_{13} \\
    \frac{1}{2} \lambda_{12} & \lambda_{2} & \frac{1}{2} \lambda_{23} \\
    \frac{1}{2} \lambda_{13} & \frac{1}{2} \lambda_{23} & \lambda_{3}
  \end{pmatrix},
  \quad
  \adj(\mathbf{\Lambda}) =
	\begin{pmatrix}
		\lambda_2 \lambda_3-\frac{1}{4}\lambda_{23}^2
		& \frac{1}{4} \lambda_{13}\lambda_{23}- \frac{1}{2} \lambda_{12}\lambda_{3}
		& \frac{1}{4} \lambda_{12}\lambda_{23}- \frac{1}{2} \lambda_{13}\lambda_{2} 
		\\
  	\frac{1}{4} \lambda_{13}\lambda_{23}- \frac{1}{2} \lambda_{12}\lambda_{3}
		& \lambda_1 \lambda_3-\frac{1}{4} \lambda_{13}^2
		& \frac{1}{4} \lambda_{12}\lambda_{13}- \frac{1}{2} \lambda_{23}\lambda_{1}
		\\
		\frac{1}{4} \lambda_{12}\lambda_{23}- \frac{1}{2} \lambda_{13}\lambda_{2}
		& \frac{1}{4} \lambda_{12}\lambda_{13}-\frac{1}{2} \lambda_{23}\lambda_{1}
		& \lambda_1 \lambda_2-\frac{1}{4}\lambda_{12}^2
	\end{pmatrix}.
\end{equation}
\end{widetext}

\begin{figure}[thb]
\begin{center}
  \includegraphics{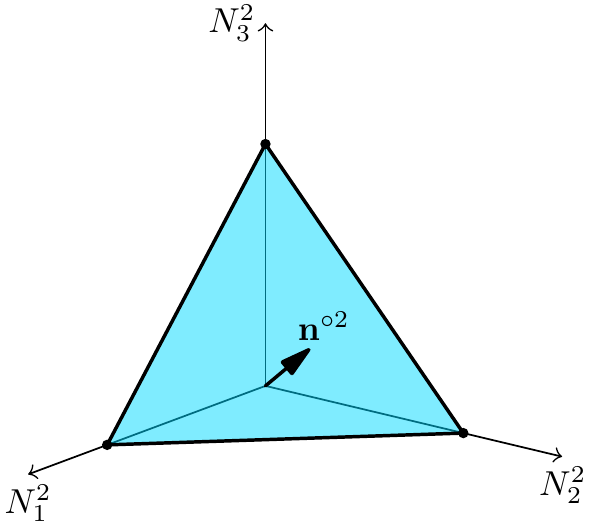}  
\caption{The unit simplex in the biquadratic three-field case with the square of the unit vector $\mathbf{n}^{\circ 2}$ along the flat direction. The flat direction can also lie along an axis or on a coordinate plane on the border of the simplex.}
\label{fig:simplex}
\end{center}
\end{figure}

The determinant of $\mathbf{\Lambda}$, which must vanish in order to have a flat direction, is \begin{equation}
\begin{split}
  4 \det(\mathbf{\Lambda}) &= 4 \lambda_{1} \lambda_{2} \lambda_{3} - \lambda_{1} \lambda_{23}^{2} 
  - \lambda_{2} \lambda_{13}^{2} - \lambda_{3} \lambda_{12}^{2} 
  \\
  &+ \lambda_{12} \lambda_{13} \lambda_{23}.
\end{split}
\end{equation}
The Hadamard square of the unit vector along the flat direction, given by Eq.~\eqref{eq:flat:direction:sol}, lies on the unit simplex illustrated in Fig. \ref{fig:simplex}. There are also six lower-dimensional solutions: the flat direction may lie on the border of the simplex on a coordinate plane or along an axis.

In spherical coordinates $N_{1} = \sin\theta \cos\phi$, $N_{2} = \sin\theta \sin\phi$, $N_{3} = \cos \theta$, the mixing angles are given by $\tan^{2} \phi = \frac{n_{2}^{2}}{n_{1}^{2}}, \quad \cos^{2} \theta = n_{3}^{2}.$

\subsection{General Two-Field Potential}

The general quartic scalar potential of two fields is 
\begin{equation}
  V= \lambda_{40} \phi_{1}^{4} + \lambda_{31} \phi_{1}^{3} \phi_{2} + \lambda_{22} \phi_{1}^{2} \phi_{2}^{2} 
  + \lambda_{13} \phi_{1} \phi_{2}^{3} + \lambda_{04} \phi_{2}^{4},
\label{eq:V:general:2:fields}
\end{equation}
where the indices of couplings count powers of the fields.

The tensor of the quartic couplings is 
\begin{equation}
  \mathbf{\Lambda} = 
  \begin{pmatrix}
    \begin{pmatrix}
      4 \lambda_{40} & \lambda_{31} \\
      \lambda_{31} & \frac{2}{3} \lambda_{22}
    \end{pmatrix}
    &
    \begin{pmatrix}
      \lambda_{31} & \frac{2}{3} \lambda_{22} \\
      \frac{2}{3} \lambda_{22} & \lambda_{13}
    \end{pmatrix}
    \\
     \begin{pmatrix}
      \lambda_{31} & \frac{2}{3} \lambda_{22} \\
      \frac{2}{3} \lambda_{22} & \lambda_{13}
    \end{pmatrix}
    &
    \begin{pmatrix}
      \frac{2}{3} \lambda_{22} & \lambda_{13} \\
      \lambda_{13} & 4 \lambda_{04}
    \end{pmatrix}
  \end{pmatrix}.
\end{equation}

The tensor eigenvalue equations are
\begin{align}
  4 \lambda_{40} N_{1}^{3} + 3 \lambda_{31} N_{1}^{2} N_{2} + 
  2 \lambda_{22} N_{1} N_{2}^{2} + \lambda_{13} N_{2}^{3} &= 4 \lambda N_{1},
  \notag
  \\
  \lambda_{31} N_{1}^{3} + 2 \lambda_{22} N_{1}^{2} N_{2} + 
  3 \lambda_{13} N_{1} N_{2}^{2} + 4 \lambda_{04} N_{2}^{3} &= 4 \lambda N_{2},
  \notag
  \\
  N_{1}^{2} + N_{2}^{2} &= 1
\label{eq:tensor:eval:eqs}
\end{align}
and the resultant of the associated homogenous equations -- the hyperdeterminant -- is
\begin{align}
  \res_{\mathbf{N}} (\mathbf{\Lambda} \mathbf{N}^{3}) 
  &= 16 [
  16 \lambda_{04} \lambda_{40} \lambda_{22}^{4} 
  - 4 (\lambda_{13}^2 \lambda_{40} + \lambda_{04} \lambda_{31}^2)  \lambda_{22}^{3}
  \notag
  \\
  & 
  - (80 \lambda_{04} \lambda_{13} \lambda_{31} \lambda_{40} + 128 \lambda_{04}^2 \lambda_{40}^2
  \notag
  \\
  & - \lambda_{13}^2 \lambda_{31}^2) \lambda_{22}^{2} + 18 (\lambda_{04} \lambda_{13} \lambda_{31}^3 
  + \lambda_{13}^3 \lambda_{31} \lambda_{40}
  \notag
  \\
  &+ 8 \lambda_{04}^2 \lambda_{31}^2 \lambda_{40}
  + 8 \lambda_{04} \lambda_{13}^2 \lambda_{40}^2) \lambda_{22} -4 \lambda_{13}^3 \lambda_{31}^3
  \notag
  \\
  &
  -27 \lambda_{04}^2 \lambda_{31}^4 -6 \lambda_{04} \lambda_{13}^2 \lambda_{31}^2 \lambda_{40}
  -27 \lambda_{13}^4 \lambda_{40}^2
  \notag
  \\
  &-192 \lambda_{04}^2 \lambda_{13} \lambda_{31} \lambda_{40}^2
  +256 \lambda_{04}^3 \lambda_{40}^3
   ].
\end{align}
We can solve $\res_{\mathbf{N}} (\mathbf{\Lambda} \mathbf{N}^{3}) = 0$ for, e.g., $\lambda_{22}$ to ensure a flat direction.

All tensor eigenvalues, satisfying Eqs.~\eqref{eq:tensor:eval:eqs}, corresponding to real tensor eigenvectors must be non-negative; more succinct criteria for the potential \eqref{eq:V:general:2:fields} to be bounded from below also exist \cite{Kannike:2016fmd}.
The Hessian matrix of the potential \eqref{eq:V:general:2:fields} must be positive-semidefinite at the extremum.

\section{Conclusions}
\label{sec:conclusions}

We propose a novel technique for investigating the appearance of a flat direction in the scalar potential of a scale-invariant model. Our method builds on the observation that, in presence of a flat direction, the determinant of the quartic coupling matrix of a biquadratic potential vanishes. This result is extended to general potentials via the formalism of tensor eigenvalues. In comparison with the usual hyperspherical coordinate approach, our matrix method noticeably simplifies the study of complicated scalar sectors, opening the way to phenomenological studies of more involved scenarios.          

\begin{acknowledgments}
We are grateful to Alexandros Karam and Antonio Racioppi for discussions. This work was supported by the Estonian Research Council grants PRG356 and PRG434, and by the EU through the ERDF CoE program project TK133.
\end{acknowledgments}

\bibliography{geom_flat}

\end{document}